\documentstyle[floats,prd,aps,epsfig,eqsecnum,12pt]{revtex}
\makeatletter
\newbox\tempboxa
\newdimen\captionboxsubcount 
\def\capsize#1{\captionboxsubcount=#1pt}
\newdimen\captionboxsub
\captionboxsub=\hsize \advance\captionboxsub by -\captionboxsubcount
\advance\captionboxsub by -\captionboxsubcount
\long\def\@makecaption#1#2{
 \setbox\@tempboxa\hbox{#1 #2}
 \ifdim \wd\@tempboxa >\captionboxsub 
\rightskip=\captionboxsubcount \leftskip=\captionboxsubcount #1 #2 
\else \hbox to\hsize{\hfil\box\@tempboxa\hfil} 
 \fi}
\makeatother
\capsize{30}

\begin{document}
\bibliographystyle{unsrt}
\begin{titlepage}
\begin{flushright}
\begin{minipage}{5cm}
\begin{flushleft}
\small
\baselineskip = 13pt
SU-4240-693\\
\end{flushleft}
\end{minipage}
\end{flushright}
\begin{center}
\Large\bf

$\eta'\rightarrow \eta\pi\pi$ Decay as a Probe of\\ a Possible
Lowest-Lying Scalar Nonet
\end{center}
\vfil
\footnotesep = 12pt
\begin{center}
\large
\quad\quad Amir H. {\sc Fariborz}
\footnote{Electronic  
address: {\tt amir@suhep.phy.syr.edu}} 
and 
Joseph {\sc Schechter}
\footnote{
Electronic address : {\tt schechte@suhep.phy.syr.edu}}\\
{\it 
\qquad  Department of Physics, Syracuse University, 
Syracuse, NY 13244-1130, USA.} \\
\vskip 0.5cm
\end{center}
\vfill
\begin{center}
\bf
Abstract
\end{center}
\begin{abstract}
\baselineskip = 17pt
We study the $\eta'\rightarrow \eta\pi\pi$ decay within an effective
chiral Lagrangian approach 
in which the lowest lying scalar  meson candidates $\sigma(560)$ and  
$\kappa(900)$ together with the $f_0(980)$ and
$a_0(980)$ are combined into a possible nonet.   
We show that there exists a unique choice of the free parameters of this
model which,  in addition to fitting the $\pi\pi$ and $\pi K$ scattering
amplitudes,  well describes the  
experimental measurements for the partial decay width of 
$\eta'\rightarrow\eta\pi\pi$ and the energy dependence of this
decay. 
  As a by-product, we  estimate the $a_0(980)$ width to be 70 MeV, 
in agreement with a new experimental analysis. 

\end{abstract}
\begin{flushleft}
\footnotesize
PACS number(s): 13.75.Lb, 11.15.Pg, 11.80.Et, 12.39.Fe 
\end{flushleft}
\vfill
\end{titlepage}
\setcounter{footnote}{0}

\section{Introduction}

Understanding  the status,  in general, and the quark content, in
particular,  of the lowest lying scalar
mesons is an issue of great current interest.   In the cases of the
$\sigma$ and
the $\kappa$ mesons, even their existence has been the subject of
many different investigations.   One may consult
refs.\cite{Torn95}-\cite{Dmi96} for
a variety of different recent works.

In the approach upon which this paper is based, a need for a $\sigma$ 
with a mass
around 560 MeV was found in the analysis of $\pi\pi$ scattering 
\cite{San95,Har96} and a need for a $\kappa$ with a mass around 900 MeV
was required in order to describe the experimental data on the  $\pi K$
scattering amplitude \cite{Blk98a}.   These investigations were carried
out in
an effective Lagrangian framework  motivated by the $1/N_c$ approximation
to QCD.
In this approach, one incorporates the contribution of 
tree Feynman diagrams,  computed from a chiral Lagrangian, including
 all possible
intermediate states within the energy region of
interest.
Furthermore, crossing symmetry is automatic, while the unknown parameters
characterizing the scalars are adjusted to satisfy the unitarity bounds.
Approximate amplitudes satisfying both crossing and unitarity are then
obtained.  For the case of $\pi K$ scattering in the $I={1\over 2}$
channel the analysis of ref.\cite{Blk98a} may be seen to be consistent
with the
experimental work of ref.\cite{Ast88}.  The experimental analysis
characterizes
the data by an effective range approximation below 1 GeV; in the treatment
of \cite{Blk98a} it is resolved into the sum of a ``current-algebra''
piece,
vector meson exchange pieces and scalar meson exchange pieces.   In
particular, the presence of a $\kappa$-meson is needed to ensure
unitarity.

  Motivated by 
the evidence for a $\sigma$ and a $\kappa$,  and taking
into account other  experimentally well-established scalars -- the
$f_0(980)$ and the $a_0(980)$ -- a possible classification of these
scalars 
(all below 1 GeV) into a nonet,
\begin{equation}
N = \left[ \begin{array}{c c c}
N_1^1&a_0^+&\kappa ^+\\
a_0^-&N_2^2&\kappa ^0\\
\kappa^-&{\bar \kappa}^0&N_3^3
\end{array} \right],
\label{N_matrix}
\end{equation}
was  studied in \cite{Blk98b}.  Since the properties of this
scalar 
nonet are expected to be less standard than those of a conventional
nonet (like the vectors), the mass piece of the effective Lagrangian is
allowed to contain extra terms:
\begin{equation}
{\cal L}_{mass} = -a {\rm Tr}(NN) - b {\rm Tr}(NN{\cal M}) - c {\rm
Tr}(N){\rm Tr}(N) - d {\rm Tr}(N) {\rm Tr}(N{\cal M}),
\label{L_mass}
\end{equation}
where ${\cal M}$ is the usual quark mass spurion.  Retaining just the $a$
and $b$  terms yields ``ideal mixing'' \cite{Oku63}.
The physical particles $\sigma$ and $f_0$ which diagonalize
the mass matrix are related to the basis states $N_3^3$ and $(N_1^1 +
N_2^2)/{\sqrt 2}$ by
\begin{equation}
\left( \begin{array}{c} \sigma\\ f_0 \end{array} \right) = \left(
\begin{array}{c c} {\rm cos} \theta_s & -{\rm sin} \theta_s \\ {\rm sin}
\theta_s & {\rm cos} \theta_s \end{array} \right) \left( \begin{array}{c}
N_3^3 \\ \frac {N_1^1 + N_2^2}{\sqrt 2} \end{array} \right),
\label{S_mix}
\end{equation}
where $\theta_s$ is the scalar mixing angle.   The coefficients $a,b,c,$
and $d$ are determined in terms of $m_\sigma$, $m_{f0}$, $m_{a0}$ and
$m_\kappa$, and for a given input set of these masses there are
two scalar mixing angles. 
Typical values of the input masses ($m_\sigma = 550$ MeV, $m_{f0}=980$
MeV, $m_{a0}=983.5$ MeV and $m_\kappa = 897$ MeV)  yield the two
possibilities:
\begin{eqnarray}
\left( a \right)  \quad &{\theta}_s& \: \approx \: -21^{\circ},
\nonumber \\
\left( b \right) \quad &{\theta}_s& \: \approx \: -89^{\circ}.
\label{theta_s}
\end{eqnarray}
In order to determine which of these two possibilities is the correct one,
it is necessary to study the pattern of scalar-pseudoscalar-pseudoscalar
interactions, which are correlated with each other by the proposed nonet
structure.
In this picture,  the general form of 
the SU(3) flavor invariant  
scalar-pseudoscalar-pseudoscalar
interaction is:
\begin{eqnarray}  
{\cal L}_{N\phi \phi} &=&
A{\epsilon}^{abc}{\epsilon}_{def}N_{a}^{d}{\partial_\mu}{\phi}_{b}^{e}
{\partial_\mu}{\phi}_{c}^{f}
+ B {\rm Tr} \left( N \right) {\rm Tr} \left({\partial_\mu}\phi
{\partial_\mu}\phi \right) + C {\rm Tr} \left( N {\partial_\mu}\phi
\right) {\rm
Tr} \left( {\partial_\mu}\phi \right) 
\nonumber \\
 &+& D {\rm Tr} \left( N \right) {\rm Tr}
\left({\partial_\mu}\phi \right)  {\rm Tr} \left( {\partial_\mu}\phi
\right),
\label{L_NPP}
\end{eqnarray}
where $\phi_a^b \left( x \right)$ is the matrix of the pseudoscalar nonet
fields, and $A, B, C, D$ are real parameters.  Derivative coupling to
the two 
pseudoscalars is used to ensure that Eq. (\ref{L_NPP}) represents the
leading
term of a chiral invariant expression (see Appendix B of \cite{Blk98b}).
It is
easy to see that all the coupling constants relevant for the study of
$\pi\pi$ and $\pi K$ scattering depend only on the parameters $A$ and $B$.
The analysis of \cite{Blk98b} then shows that possibility (a) in
(\ref{theta_s}) for
the scalar mixing angle is selected as the correct one in the present
scheme.   The parameters $C$ and $D$ were left undetermined in the
analysis of \cite{Blk98b}, as no scalar-pseudoscalar-pseudoscalar coupling
involving an $\eta$ or $\eta'$ was present in the $\pi\pi$ and $\pi K$
scattering discussed there. 

In this work we explore the parameter space of $C$ and $D$  in 
detail by studying the $\eta'\rightarrow \eta\pi\pi$ decay,  for which
there are relatively 
recent and
precise experimental measurements.
As we will see, the scalar couplings to $\eta$ and $\eta'$ play
a dominant role in the amplitude for this decay.

All the discussion in the present paper will use the same methods and
parameters as in the previous $\pi\pi$ and $\pi K$ scattering papers
\cite{Har96,Blk98a}.   Thus, this work can be thought of as a check of
that method as
well as a test of the basic assumption that the low lying scalars are
related to each other by belonging to a (broken) flavor SU(3) nonet.  In
the sense that the effective Lagrangian method makes no explicit reference
to the quark structure  of these scalars, the present work may be
considered model independent.  Note also that only the SU(3) flavor
structure of the scalars is required to construct non-linear chiral
Lagrangians describing these interactions \cite{Cal69}.

``Microscopic'' models of low lying scalars have been suggested in
which they are variously $ qq {\bar q}{\bar q}$ states in the MIT bag
\cite{Jaf77}, meson-meson molecules \cite{Isg90} or unitarity corrections
due to
strong meson meson interactions \cite{Torn95,Ish98}.  All these models
involve
four quarks and so may be related to each other.  A ``model-independent''
effective Lagrangian might be an appropriate vehicle for summarizing the
common feature of different microscopic models.

The process $\eta'\rightarrow\eta \pi\pi$ has been studied by many
authors in chiral symmetric frameworks since the early days of
``Current-Algebra''.  Treatments have used exclusively contact terms
\cite{Cro67,Schw68,Maj68,DiV81} or contact terms plus scalar meson
exchanges \cite{Sch73,Sing75,Desh78,Bra80}. Ordinarily
in the chiral perturbation theory approach \cite{Wein79} all effects of
resonance exchanges are assumed to be ``integrated out'' and summarized in
the complete set of contact terms.     However, in the case of the
$\eta'(958)$ decay, the masses of the intermediate $\sigma$, $f_0$ and
$a_0$ resonances are either less than or comparable to 958 MeV.  Thus,
kinematical dependences due to the propagators could be important.  The
new features of the present treatment include the use of
Eq.(\ref{L_mass})
 to describe the scalar mesons and mixing angle, the use of
Eq.(\ref{L_NPP})
to describe the scalar coupling constants and a procedure uniform with the   
discussion of $\pi\pi$ and $\pi K$ scattering in
\cite{San95,Har96,Blk98a}. Furthermore,
comparison is being made with more recent data.

This paper is organized as follows.  Section II gives our theoretical
prediction of the $\eta'\rightarrow \eta\pi\pi$ process as well as the
experimental parameterization.  The fit to the experiment, taking into
account the experimental uncertainties, is treated in detail in Section
III.   Finally, Section IV continue a brief summary and discussion.


\section{$\eta'\rightarrow \eta\pi\pi$ decay}

Here, we will predict the amplitude for this process in the present model
and display the experimental data to which it will be compared.

\begin{figure}
\centering
\epsfig{file=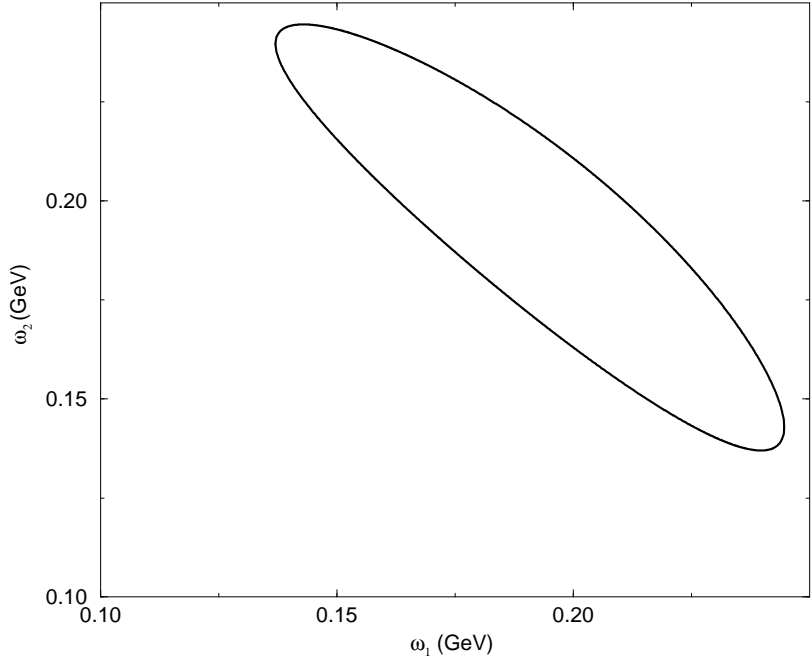, height=5in, angle=270}
\caption
{
The boundary of integration in Eq.(\ref{etap_dw}).
}
\label{Fig_w1w2}
\end{figure}

We assume exact iso-spin invariance which seems consistent with the
present experimental accuracy.  The four momenta of the particles are
labeled according to the scheme $\eta'(p)\rightarrow \eta(k) +\pi_1(q_1)
+ \pi_2(q_2)$, wherein ($\pi_1, \pi_2$) can stand for either ($\pi^+,
\pi^-$) or ($\pi^0, \pi^0$).   The partial widths are related to the
invariant matrix element $M(p\rightarrow k + q_1 + q_2)$ by

\begin{equation}
\Gamma(\eta'\rightarrow\eta\pi^+\pi^-)=
2\Gamma(\eta'\rightarrow\eta\pi^0\pi^0)=
{1\over {2 m_{\eta'}}}\int|M|^2d\Phi,
\label{etap_width}
\end{equation}
where the phase space volume element $d\Phi$ is
\begin{equation}
d\Phi=(2\pi)^4 \delta^4(p-k-q_1-q_2)
{ {d{\bf k}}\over {2\omega (2\pi)^3} }
{ {d{\bf q_1}}\over {2\omega_1 (2\pi)^3} }
{ {d{\bf q_2}}\over {2\omega_2 (2\pi)^3} },
\end{equation}
with $\omega = \sqrt{m_\eta^2 + {\bf k}^2 }
$ and  $\omega_i = \sqrt{m_{\pi}^2 + {\bf q_i}^2 } $.
After performing the usual phase space integration we have
\begin{equation}
\Gamma_{\eta'\rightarrow\eta\pi\pi}=
{1\over {64\pi^3 m_\eta'}}
\int d{\omega_1}d{\omega_2}
\left| M \right|^2.
\label{etap_dw}
\end{equation}
The boundary of integration in the $\omega_1\omega_2$ plane for our choice
of  $m_\pi=137$ MeV, $m_\eta=547$ MeV and $m_{\eta'}=958$ MeV \cite{PDG} 
is shown in Fig.\ref{Fig_w1w2}.

In the treatment of $\pi\pi$ \cite{San95,Har96} and $\pi K$
\cite{Blk98a}
scattering
according to the present approach it was found that a reasonable
approximation up to the 1 GeV energy range consisted of  including i) the
``current algebra'' contact term ii) vector meson tree diagrams and iii)
light scalar ($f_0(980), \sigma, \kappa$) meson tree diagrams.   These
were all calculated from a chiral Lagrangian with the minimum number of
derivatives.  For $\eta'\rightarrow\eta\pi\pi$ there is a big
simplification since G-parity conservation shows that no vector meson
exchanges are possible at tree level.  Similarly, the derivative part of
the contact term vanishes.

\begin{figure}
\centering
\epsfig{file=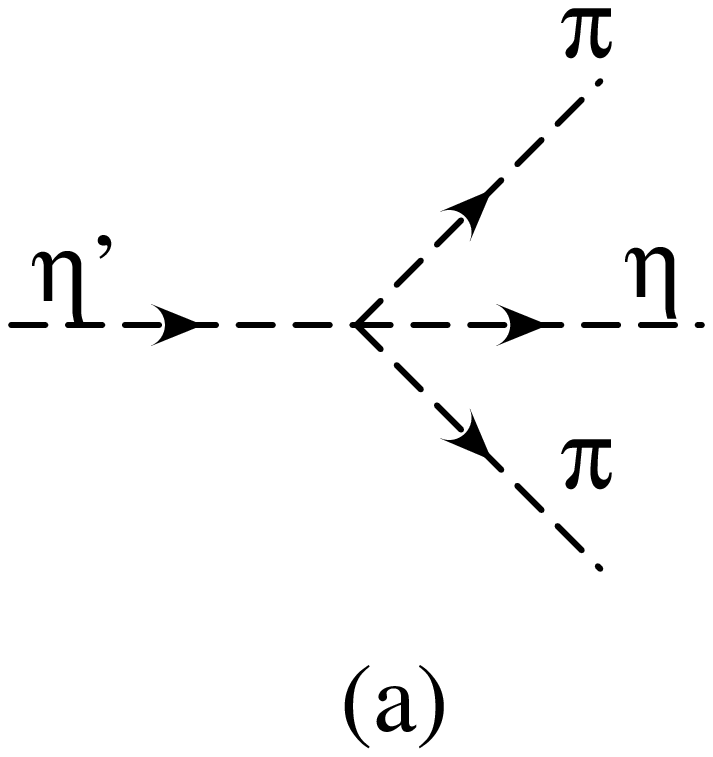, height=2in, angle=0}

\epsfig{file=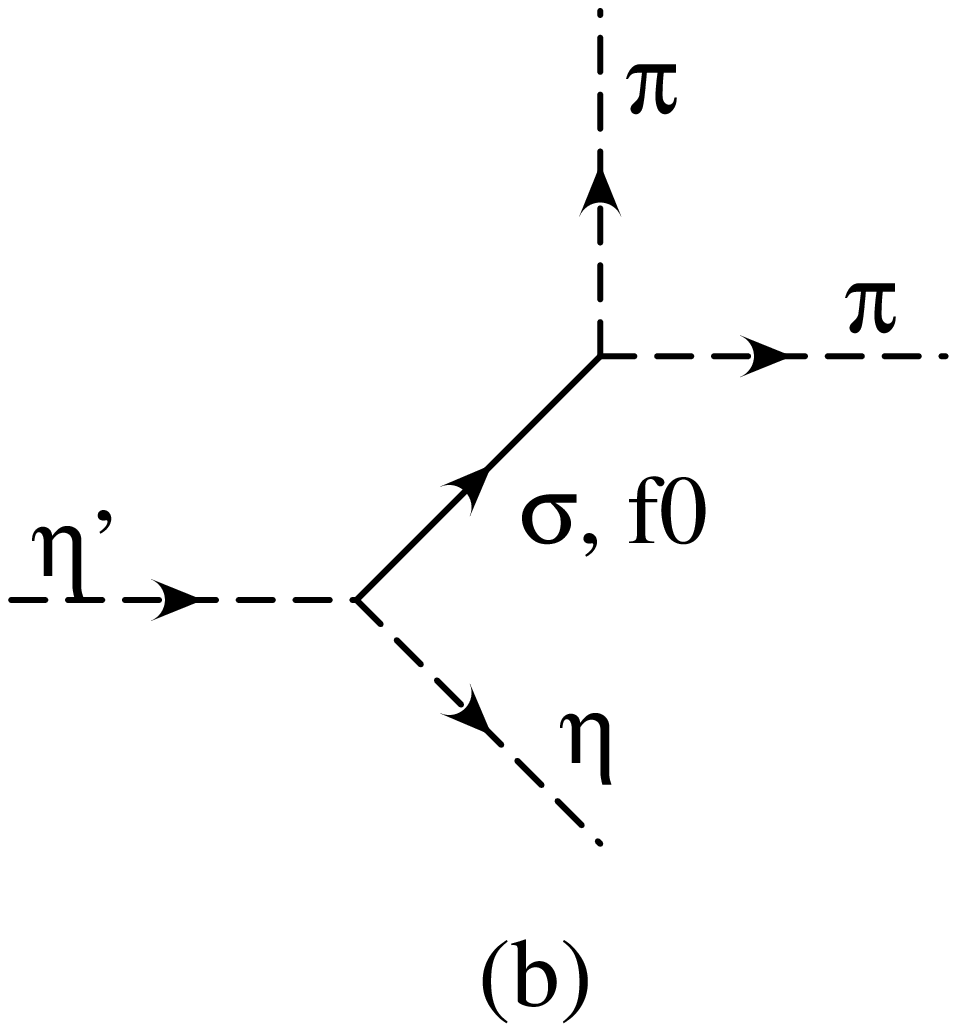, height=2.5in, angle=0}

\epsfig{file=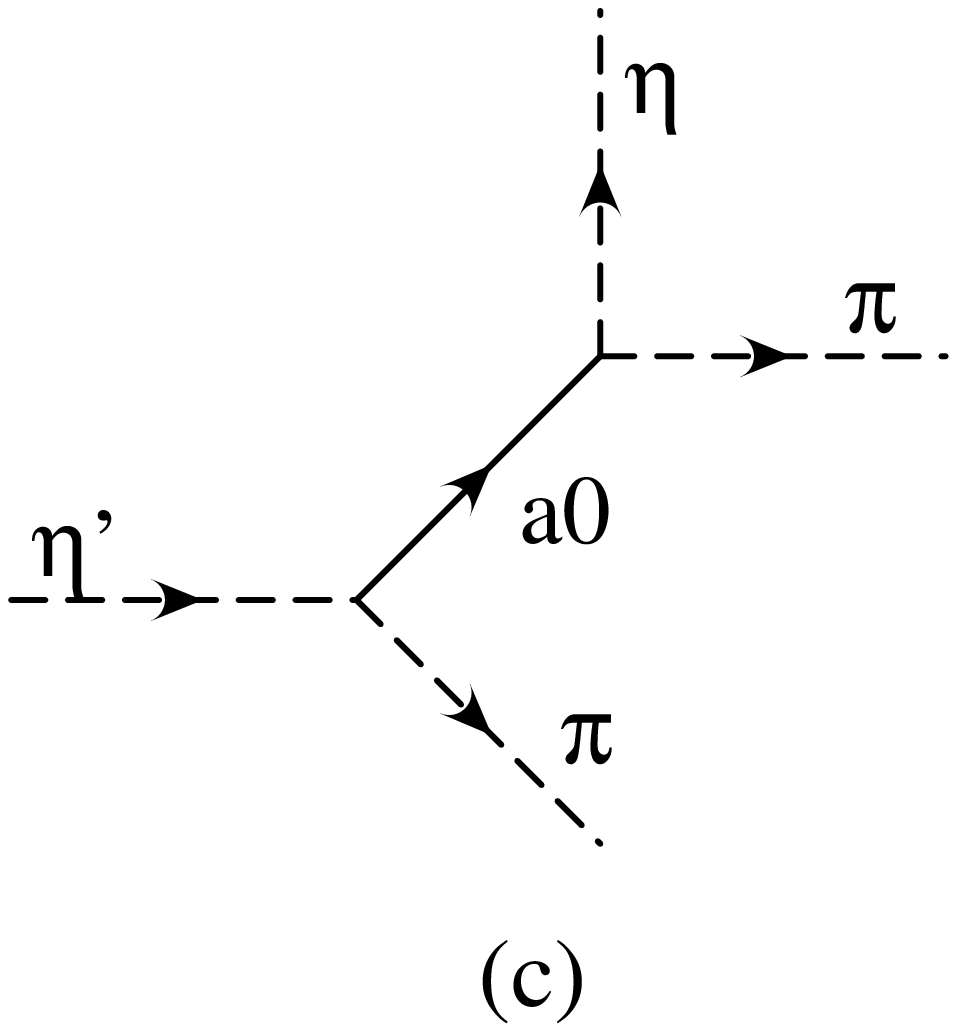, height=2.5in, angle=0}

\caption
{
Tree Feynman diagrams  representing the contributions of (a) the
current algebra, (b) the $\sigma$ and the $f_0(980)$,  and (c) 
the $a_0(980)$ terms to the decay
$\eta'\rightarrow\eta\pi\pi$ in our model. }
\label{FD}
\end{figure}

The individual contributions shown in Fig.\ref{FD} are then
\begin{eqnarray}
M_{C.A.}&=& {  {m_\pi^2}  \over {F_\pi^2} }  {\rm sin}2\theta_p,
\nonumber \\
M_\sigma &=& -\sqrt{2}\gamma_{\sigma\eta\eta'}\gamma_{\sigma\pi\pi}
{
  {(p.k)(q_1.q_2)}
    \over
  { m_\sigma^2 + (p-k)^2 -i m_\sigma G'_\sigma}
},
\nonumber \\
M_{f_0} &=& -\sqrt{2}\gamma_{f\eta\eta'}\gamma_{f\pi\pi}
{
  {(p.k)(q_1.q_2)}
    \over
  { m_{f_0}^2 + (p-k)^2 - im_{f0}G'_{f0} }
},
\nonumber \\
M_{a_0} &=&  -\gamma_{a\pi\eta'}\gamma_{a\pi\eta}
\left[
{
  {(p.q_2)(k.q_1)}
    \over
  { m_{a_0}^2 + (p-q_2)^2 -i m_{a0} G'_{a0}}
}
+
{
  {(p.q_1)(k.q_2)} 
    \over
  { m_{a_0}^2 + (p-q_1)^2 -i m_{a0} G'_{a0} }
}
\right].  
\label{M_individual}
\end{eqnarray}
 
The total decay amplitude $M$ is the sum of these pieces.  The current
algebra contribution $M_{C.A.}$ is obtained from the ``quark mass'' term
in the effective Lagrangian (proportional to ${\rm Tr} \left[ (U +
U^\dagger) {\cal M} \right]$, where $U={\rm exp} \left[ 2i\phi
/F_\pi\right]$).   Definitions of the various
scalar-pseudoscalar-pseudoscalar coupling constants which appear in the
$\sigma, f_0(980)$ and $a_0(980)$ exchange diagrams are given in Appendix
A. These involve the coefficients $A,B,C,D$ of Eq. (\ref{L_mass});$A$ and
$B$
were previously found from $\pi\pi$ and $\pi K$ scattering while $C$ and
$D$ remain to be determined here.  The scalar masses are taken as
mentioned before Eq.(\ref{theta_s}).
Even though there is no kinematical possibility
for any of the intermediate scalars to be on the ``mass shell'' we include
``total width'' terms in the propagator denominators in order to agree
with the previous work \cite{San95,Har96,Blk98a}.
The $f_0$ and $a_0$ exchange terms will be essentially taken to be 
of Breit-Wigner form so $G'_{f0}$ and $G'_{a0}$ are related to the
coupling 
constants.  We take $G'_{f0}=64.6$ MeV from \cite{Har96} (\cite{PDG}
allows
40--100 MeV) and $G'_{a0}=50-100$ MeV \cite{PDG}.  The exact value of
$G'_{a0}$ will be found from our analysis since it depends on the
parameter $C$.   Finally we take $G'_\sigma=370$ MeV \cite{PDG}; 
this is related to a pole position rather than a total Breit-Wigner width,
a prescription which enables the construction of a $\pi\pi$
amplitude satisfying both the unitarity bounds and crossing symmetry.

The theoretical expressions in Eq. (\ref{etap_width})-(\ref{M_individual})
will be compared with
the experimental data on partial decay rates and energy dependence of
$|M^2|$.  The experimental results for the rates are listed \cite{PDG}
as:
\begin{eqnarray}
\Gamma^{exp}_{\eta'\rightarrow\eta\pi^+\pi^-}
&=& 0.089\pm 0.010 \hskip .15cm {\rm MeV}
\nonumber \\
2\Gamma^{exp}_{\eta'\rightarrow\eta\pi^0\pi^0}
&=& 0.084 \pm 0.012 \hskip .15cm {\rm MeV} 
\end{eqnarray}
in agreement with iso-spin invariance.  Since we are working in the
iso-spin invariant limit we will average 
\footnote{
For the average value ${\bar x} +\delta{\bar x}$ of measurements  $x_i
+\delta
x_i$, we use ${\bar x}=\sum_i x_i w_i/\sum_i w_i ; \delta {\bar x}=
(\sum_i w_i)^{-1/2}$ with the weight $w_i=1/(\delta x_i)^2$.
}    
these to obtain:
\begin{equation}
\Gamma^{exp}_{\eta'\rightarrow\eta\pi\pi} =
0.0872 \pm 0.008 \hskip .15cm {\rm MeV},
\label{etap_exp}
\end{equation}
with which the theoretical results will be compared.

For describing the energy dependence, experimentalists use the Dalitz-like
variables\cite{Ald86}:

\begin{eqnarray}
x &=& {\sqrt{3}\over Q} (\omega_1 - \omega_2)
\nonumber \\
y &=& -{ {2+ m_\eta /m_\pi } \over Q } (\omega_1 + \omega_2 ) -1
+ { { 2 + m_\eta / m_\pi }\over Q }(m_{\eta'} - m_\eta )
\end{eqnarray}
with $Q=m_{\eta'}-m_\eta - 2 m_\pi$.   As $\omega_1$ and $\omega_2$
vary over the physical region in Fig.\ref{Fig_w1w2}, $x$ ranges from about
-1.4
to 1.4 and $y$ ranges from -1 to about  1.2.  One may expand the
matrix
element, up to an irrelevant overall phase, as
\begin{equation}
M={\cal A}^{1/2} \left[ 1 + \beta_1 y + \beta_2 y^2 + \gamma_2 x^2\right]
+ \cdots
\end{equation}
where ${\cal A}$ is real while $\beta_1$, $\beta_2$ and $\gamma_2$ are
complex. 
The expansion begins with $x^2$ since $M$ (see for example
Eq.(\ref{M_individual}))
must be invariant on the interchange $q_1\leftrightarrow q_2$, which
implies $x
\leftrightarrow - x$.
 It is found \cite{Ald86} that this form yields an $M^2$ which fits
the experimental data when the $y x^2$, $y^3$, $x^4$ and $y^2 x^2$ terms
are negligible:
\begin{equation}
|M|^2={\cal A} \left[     \left| 1 + \alpha y \right|^2 
                          + {\tilde c} x^2 
               \right] 
               + \cdots
\label{M_expand}
\end{equation}
Here $\alpha$ is complex and ${\tilde c}$ is real. 
For the decay $\eta'\rightarrow \eta\pi^0\pi^0$, the experimental
values  are \cite{Ald86}  
\begin{eqnarray}
{\rm Re}\hskip .15cm \alpha &=& -0.058 \pm 0.013 
\nonumber \\
{\rm Im}\hskip .15cm \alpha &=& 0.00\pm 0.13
\nonumber \\
{\tilde c} &=& 0.00 \pm 0.03, 
\end{eqnarray}
and for the  decay $\eta'\rightarrow \eta\pi^+\pi^-$ 
\begin{equation}
{\rm Re}\hskip .15cm \alpha=-0.08\pm 0.03.
\end{equation}
As explained before,  we  compare our results with the
average of the experimental data for charged and neutral pions.
This means  we should match our results to 
\begin{eqnarray}
{\rm Re}\hskip .15cm \alpha &=& -0.062 \pm  0.012
\nonumber \\
{\rm Im}\hskip .15cm \alpha &=& 0.00\pm 0.13
\nonumber \\
{\tilde c} &=& 0.00 \pm 0.03.
\label{ac_exp}
\end{eqnarray}
The parameter ${\cal A}$ in Eq.(\ref{M_expand}) is determined using Eq.
(\ref{etap_exp}).

Altogether, the experimental data are fit with the four real quantities
${\cal A}$, Re $\alpha$, Im $\alpha$ and ${\tilde c}$.
On the other hand the theoretical expression in Eq.(\ref{M_individual}) is
completely
fixed if we specify just the two real constants $C$ and $D$ in Eq.
(\ref{L_mass}), since everything else is already specified.  Clearly there
is no
{\it a priori} guarantee that we can fit the data using the present model.
Furthermore,  it is necessary for the expansion of Eq.(\ref{M_individual})
to also
yield negligible higher order terms in Eq.(\ref{M_expand}).   We will see
in the
next section that there in fact exists a unique choice of $C$ and $D$
which can fit the experimental data.


\section{Fit to Experiment}

Our job is to find the parameters $C$ and $D$ so that $|M|^2$ computed
from Eq.(\ref{M_individual}) agree with the experimental form given in
Eqs. (\ref{M_expand})
, (\ref{etap_exp}) and (\ref{ac_exp}) up to the stated uncertainties.

As a preliminary we note that restrictions on the allowed
values of $C$ may be obtained from experimental information on 
$a_0(980)\rightarrow \pi\eta$ decay.  This partial width is given by
\begin{equation}
\Gamma (a_0\rightarrow\pi\eta)=
{ \gamma_{a\pi\eta}^2 q\over {32\pi m_{a0}^2 }  }
\left( m_{a0}^2 -m_\pi^2 -m_\eta^2\right)
\end{equation}
where $q$ is the center of mass momentum of the final state mesons.
Now Eq.(\ref{gamma_ape}) of Appendix A shows that $\gamma_{a\pi\eta}$
depends on the known values of $A$ and $\theta_p$ as well as the unknown 
value of $C$.  The Review of Particle Properties \cite{PDG} lists the
total $a_0$ width as 50--100 MeV and the $\pi\eta$ mode as ``dominant''.
It was estimated in the present model (section IV of \cite{Blk98b}) that
$\Gamma
(a_0\rightarrow K{\bar K})$ is only about 5 MeV so we expect 
$G'_{a0} \approx \Gamma(a_0\rightarrow \pi\eta )$ +  5 MeV.   We
conservatively expect $\Gamma (a_0\rightarrow \pi\eta )$ to lie in the
range 25--100 MeV.  This restricts $C$ to the two intervals [-21, -13]
GeV$^{-1}$ and [2, 10.5] GeV$^{-1}$.

\begin{figure}
\centering
\epsfig{file=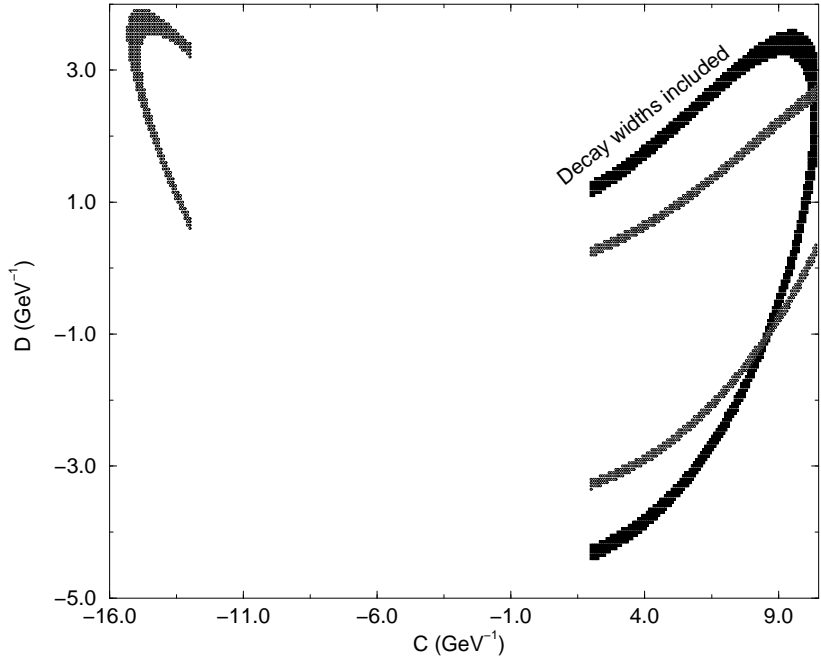, height=5in, angle=270}
\caption
{
Regions consistent with the partial decay width of
$\eta'\rightarrow\eta\pi\pi$ and $a_0(980)\rightarrow\pi\eta$.  The
semi-closed region on the right is
obtained by inclusion of
the decay widths in the propagators of the intermediate scalars.
$G'_\sigma=370$ MeV, $G'_{f0}=64.4$ MeV and $G'_{a0}=100$ MeV.
The other regions correspond to neglecting the decay widths.
}
\label{Fig_CDb} 
\end{figure}

\begin{figure}
\centering
\epsfig{file=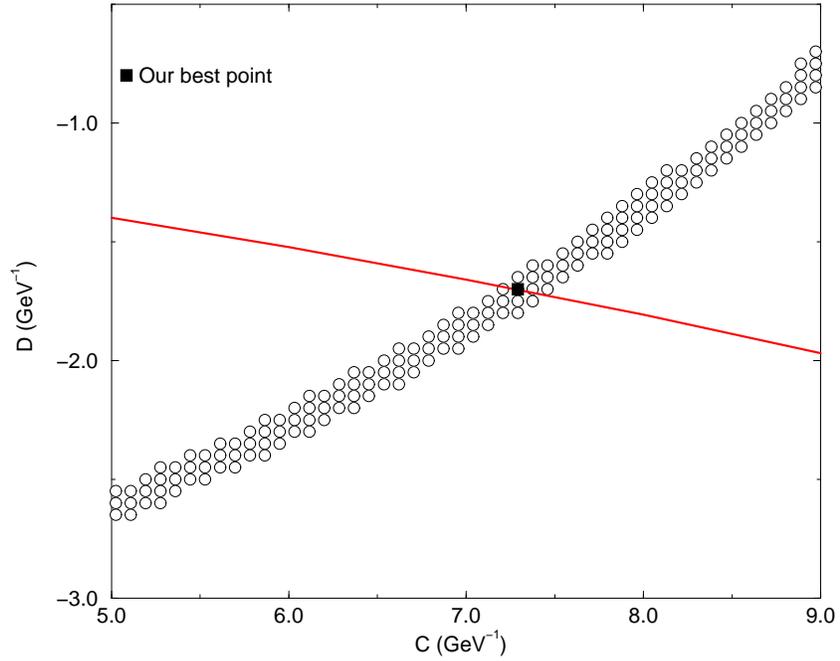, height=5in, angle=270}
\caption
{
Extracting $C$ and $D$ from two different experimental measurements on 
$\eta'$ decay.
Circles represent the region consistent with the partial decay
width of $\eta'$, and the solid line represents the least squared fits 
of the normalized magnitude squared of the decay matrix element to the
form $(1 +\alpha y)^2 + {\tilde c} x^2$ with $\alpha =-0.0615$. 
}
\label{Fig_CD_alpha}
\end{figure}

For initial orientation we shall neglect the imaginary terms in the 
denominators of Eq.(\ref{M_individual}).  We start by numerically 
\footnote{In our computation we choose $\theta_s = -20.33^{\circ}$.}
scanning the above 
two intervals of $C$ and 
searching  for the acceptable regions in the $CD$ plane that are 
consistent with
the averaged experimental partial decay width (\ref{etap_exp}).
The result of this search is shown in Fig. \ref{Fig_CDb} which also shows
the analogous intervals when the imaginary terms in Eq.(\ref{M_individual})
are
retained. 
For $C$ in the interval [-13, -21] GeV$^{-1}$  there is a small acceptable
region,
whereas for $C$ in [2, 10.5] GeV$^{-1}$  there are two acceptable regions
in the 
form of strips along the $C$ axis. In both intervals the thickness of
these
regions is  related  to the error in the averaged experimental partial
decay width 
in (\ref{etap_exp}), and therefore,  is the main source of our error
estimation in the final evaluation of $C$ and $D$.   It turns out that it
is a reasonable approximation to neglect the additional uncertainty
associated with the stated error in Re $\alpha$.
In order to further restrict the acceptable values of $C$ and $D$, we
compare our predicted energy dependence, 
$
  {\left| M(x,y) \right|^2 }
 /   {\left| M(0,0) \right|^2 }
$, with the experimental result (\ref{M_expand}) and (\ref{ac_exp}) taking 
Im $\alpha\equiv 0$ for now and ${\tilde c}$ as a fitting parameter.
We find that only the region around $C=7$ with negative $D$ 
has the required property and therefore we are left with the lower
strip in Fig.\ref{Fig_CDb}.    In Fig.\ref{Fig_CD_alpha} this region is
enlarged; 
also shown is the line representing a set of ``least squared'' minima on
which $\alpha$ is fixed.  For a given $C$, the corresponding minimum is
obtained by varying $D$ and ${\tilde c}$.  The intersection of this line
with the previous region yields the desired $C$ and $D$ estimates.  Note
that the fit improves in the direction of increasing $C$.  The values of
$C$ and $D$ are displayed in the first column of Table \ref{T_I}.

\begin{table}
\begin{center}
$\begin{array}{|c||c|c|c|}
\hline 
A ({\rm GeV}^{-1})&      2.51       &     2.51        &   2.87  
\\ \hline
B ({\rm GeV}^{-1}) &     -1.95       &    -1.95        &   -2.34
\\ \hline
C ({\rm GeV}^{-1}) &  7.29 \pm 0.08  &  7.16\pm 0.13   &  7.25\pm 0.10  
\\ \hline
D ({\rm GeV}^{-1}) & -1.70 \pm 0.08  & -2.26\pm 0.13   &  -2.09\pm 0.10
\\ \hline
{\rm Im}\hskip .15cm \alpha &        0        &  -0.12\pm 0.27  & -0.16\pm
0.20
\\ \hline
{\tilde c}     & -0.004 \pm 0.031& -0.014\pm 0.033 & -0.013\pm 0.033    
\\ \hline
\Sigma    &  1.49 & 0.0032  & 0.0045 
\\ \hline
\end{array}$
\end{center}  
\caption{
Extracted parameters from a fit of the normalized magnitude of the 
$\eta'$ decay matrix element to the form $(1+\alpha y)^2 +c x^2$, with
 Re $\alpha=-0.0615$. In the first and second columns  $m_\kappa=897$
MeV while in the last column $m_\kappa=875$ MeV.
The imaginary terms in the propagator denominators were not included 
for column 1. $\Sigma$ is the least square deviation with 1701 data
points measuring the goodness of fit.}
\label{T_I}
\end{table}

\begin{figure}
\centering
\epsfig{file=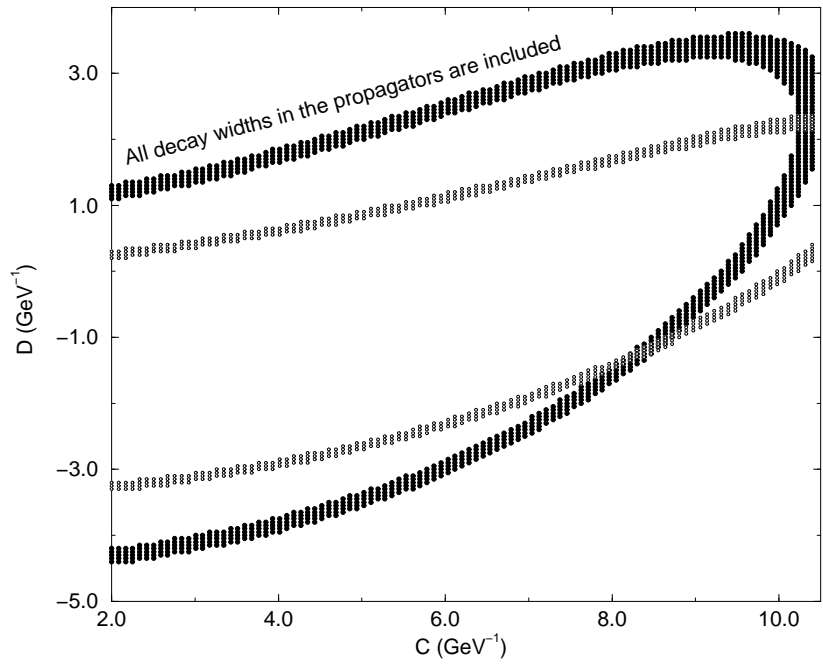, height=5in, angle=270}
\caption
{
The effect of including the widths in the propagators is dominated
by $G'_\sigma$.  In the two parallel  regions in the middle,   $G'_\sigma$
is
removed from its propagator.}
\label{Fig_CD_widths}
\end{figure}

\begin{figure}
\centering
\epsfig{file=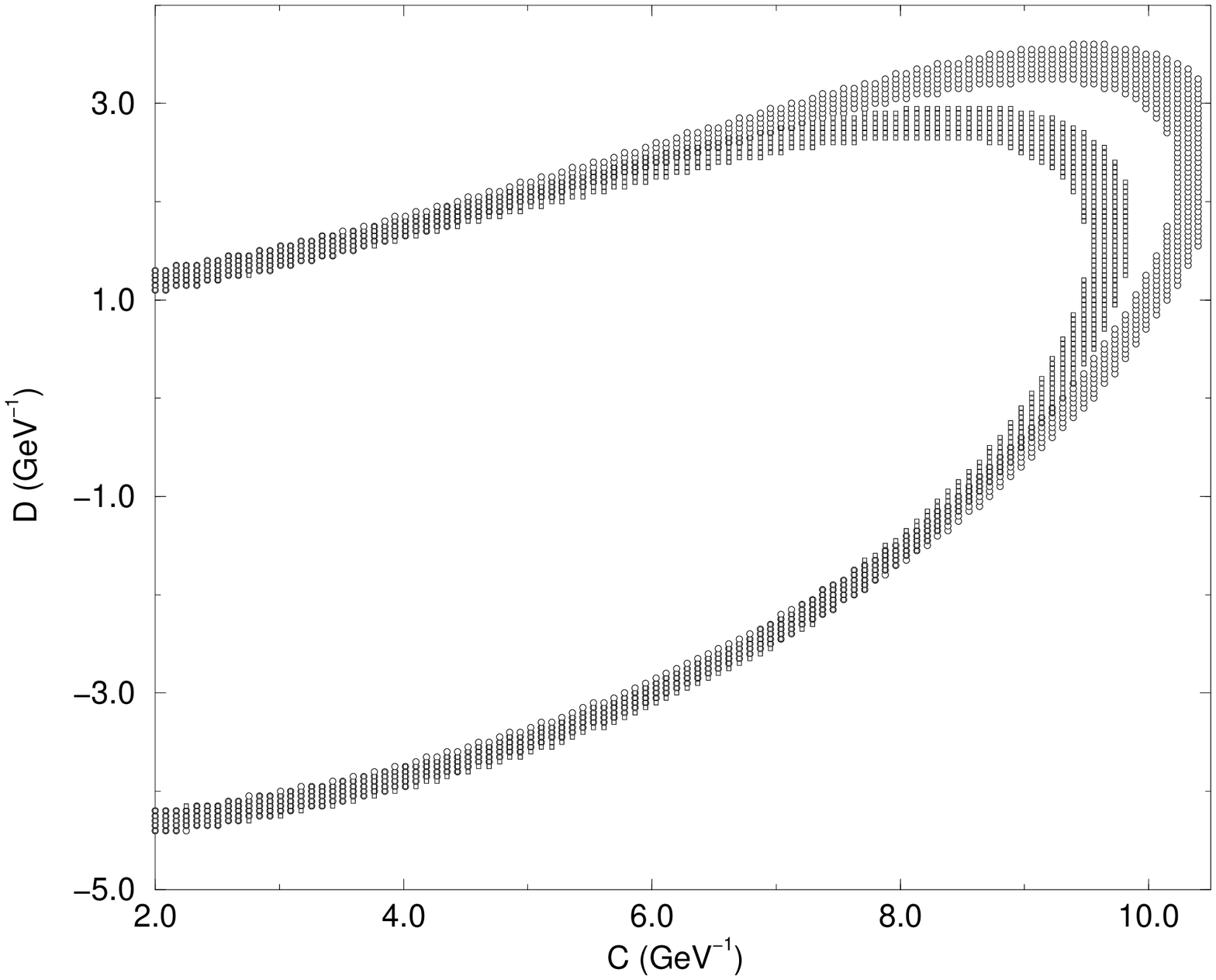, height=5in, angle=270}
\caption
{
The available region  consistent with the partial decay width
of
$\eta'\rightarrow\eta\pi\pi$ is not sensitive to $G'_{a0}$
in the physical region of $C\approx 7$.  The outer/inner regions are
obtained  with  $G'_{a0}$=100/50 MeV.  
}
\label{Fig_CD_Gpa0}
\end{figure}

Now let us include the imaginary terms in the denominators of
Eq.(\ref{M_individual}).
In our computation we choose $G'_\sigma = 370$ MeV and
$G'_{f0}=64.6$ MeV  as were obtained in \cite{Har96}, and the  two extreme
possibilities  $G'_{a0}=50, 100$ MeV.  We rescan the $CD$ plane for 
regions that are consistent with the partial decay width (\ref{etap_exp}).
The result is shown in Fig.\ref{Fig_CDb} and is also compared with
the previous
case where no widths were included.   This figure shows that
in the new case,  there is no available region for $C$ in the interval
[-21, -13] GeV$^{-1}$.   
For $C$ in the interval [2, 10.5] GeV$^{-1}$, we have shown in 
Fig.\ref{Fig_CD_widths} that the main effect of the inclusion of 
the decay widths is driven by the $\sigma$ width. 
In Fig.\ref{Fig_CD_Gpa0}, we have shown that the uncertainty in 
$G'_{a0}$
does not make a substantial difference, in particular in the physical 
region where $C \approx 7$.

\begin{figure}
\centering
\epsfig{file=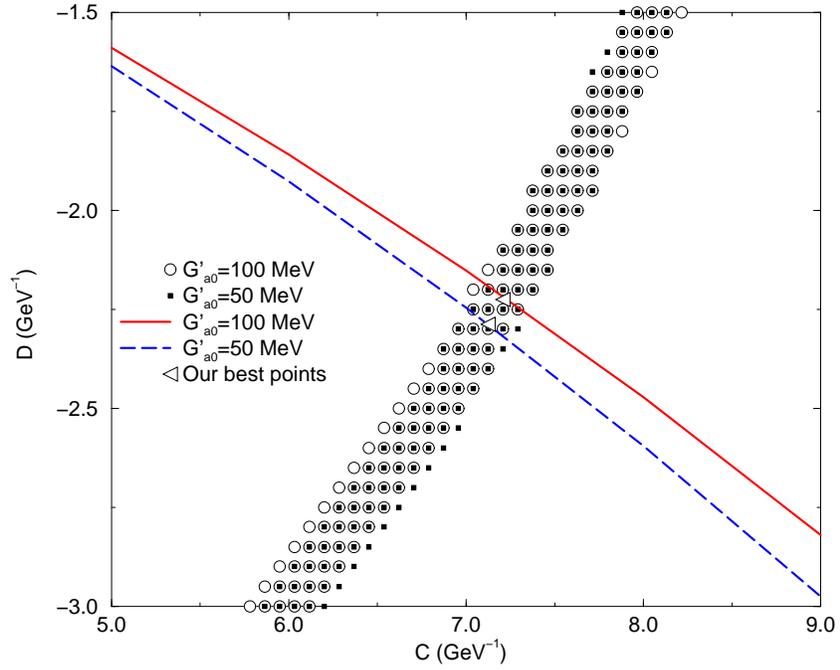, height=5in, angle=270}
\caption
{
Extracting $C$ and $D$ from two different experimental measurements on 
$\eta'$ decay.
Circles represent the region which is consistent with the partial decay
width of $\eta'$, and lines represent the least squared fits 
of the normalized magnitude of decay matrix element to the form 
$\left| 1 +\alpha y \right|^2 + {\tilde c}x^2$ with  Re $\alpha=-0.0615$,
$G'_\sigma=370$ MeV and  $G'_{f0}=64.6$ MeV. 
}
\label{Fig_CD_alphab}
\end{figure}

We proceed as before,  further restricting the available regions in the
$CD$ plane  by fitting the normalized magnitude of the decay matrix
element $M$ to the form (\ref{M_expand}) with complex $\alpha$.
We set 
${\rm Re}\alpha=-0.0615$ and fit for ${\rm Im} \alpha$ and ${\tilde c}$
in this region.  We find that the acceptable region in this case is
very close to the previous  region in Fig.\ref{Fig_CD_alpha}. The
result is
shown in Fig.\ref{Fig_CD_alphab}. The two  lines 
correspond to two values of $G'_{a0}$, and their intersections with 
the acceptable region for   $\eta'$ partial decay width provide our best
points in this plane. We however notice
that $C$ and $D$ for the value of $G'_{a0}=50$ MeV correspond to a value
of $\Gamma (a_0(980)\rightarrow\pi\eta) \approx 64$ MeV which is greater
than the total decay width itself and cannot be correct.   This
consistency  check within our computation further restricts the
experimentally  unknown value of $G'_{a0}$.  On the other hand,  the
intersection of  the line corresponding to $G'_{a0}=100$ MeV with the
acceptable region
of  $\eta'$ partial decay width gives $\Gamma_{a0(980)\rightarrow\pi\eta}
\approx 65$ MeV.  Therefore we conclude that our computation provides a
stable estimate of the partial decay width of  
$a0(980)\rightarrow\pi\eta$ to be approximately 65 MeV.  

The only other hadronic $a_0$ decay mode which has been observed
\cite{PDG} is $K{\bar K}$; using $\Gamma (a_0\rightarrow K {\bar K} )
\approx 5$ MeV \cite{Blk98b} we get an estimate $\Gamma_{tot}(a_0)\approx
70$ 
MeV.   The extracted values of $C$ and $D$ and other fitting parameters
are listed in the second column of Table \ref{T_I}.  Note
that the
goodness
of fit improves appreciably when we allow for non-zero widths.

\begin{figure}  
\centering
\epsfig{file=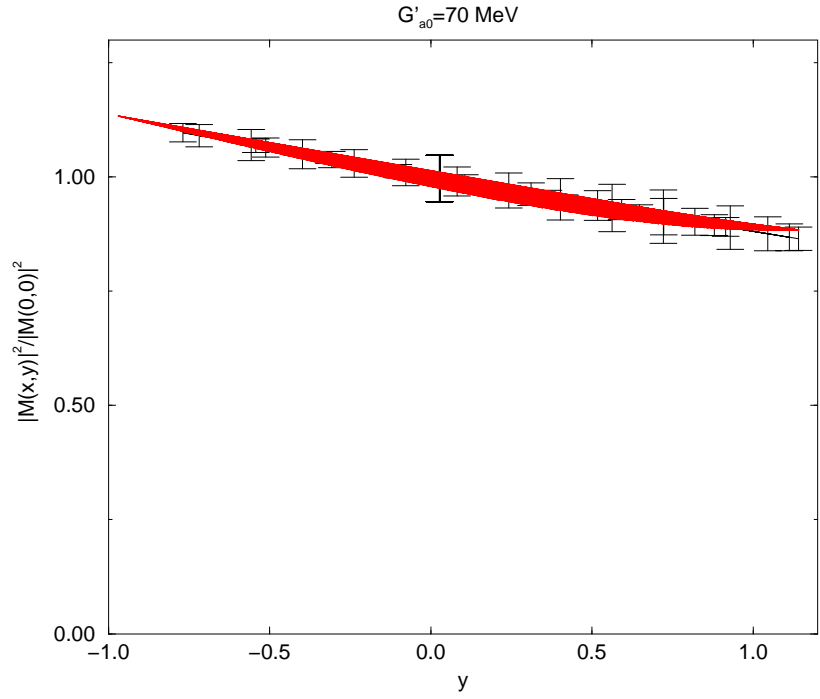, height=5in, angle=270}

\epsfig{file=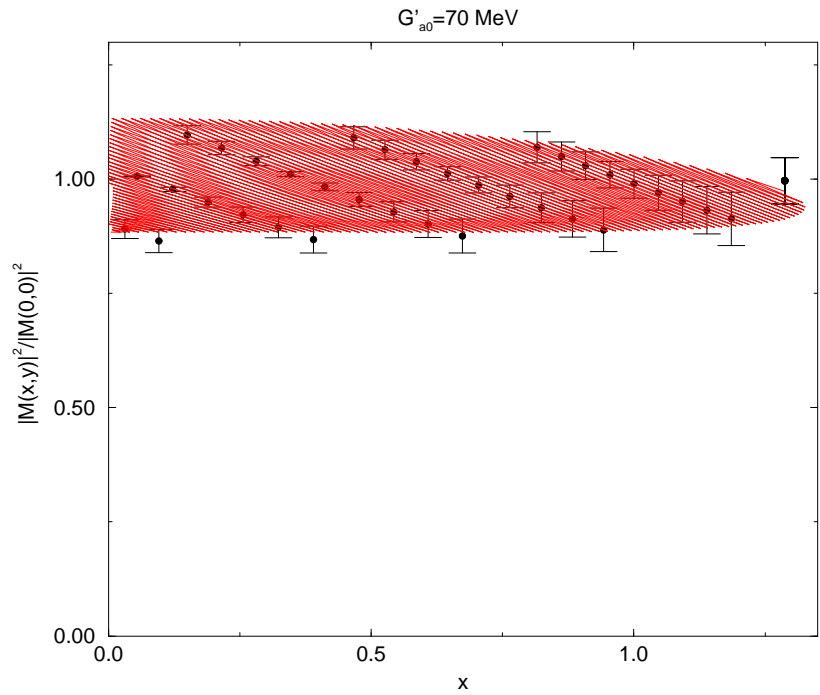, height=5in, angle=270}
\caption
{
Projections of $|{\hat M}|^2=|M(x,y)|^2/|M(0,0)|^2$ onto the 
$y-|{\hat M}^2|$ and $x-|{\hat M}^2|$ planes. Parameters as in the second
column of Table \ref{T_I}.
}
\label{Fig_M2_vs_xy}
\end{figure}

It is perhaps interesting to display the $x$ and $y$ dependences of our 
normalized matrix element squared 
$|{\hat M}|^2 = {  { | M(x,y) |^2 } / { | M(0,0) |^2 } }$.  In
Fig.\ref{Fig_M2_vs_xy} we
show the
projections of this two dimensional surface onto the $y-|{\hat M}|^2$ and 
$x-|{\hat M}|^2$ planes.  It is clear from the $y-|{\hat M}|^2$ projection
that $|{\hat M}|^2$ has very little dependence on $x$.

\begin{figure}
\centering
\epsfig{file=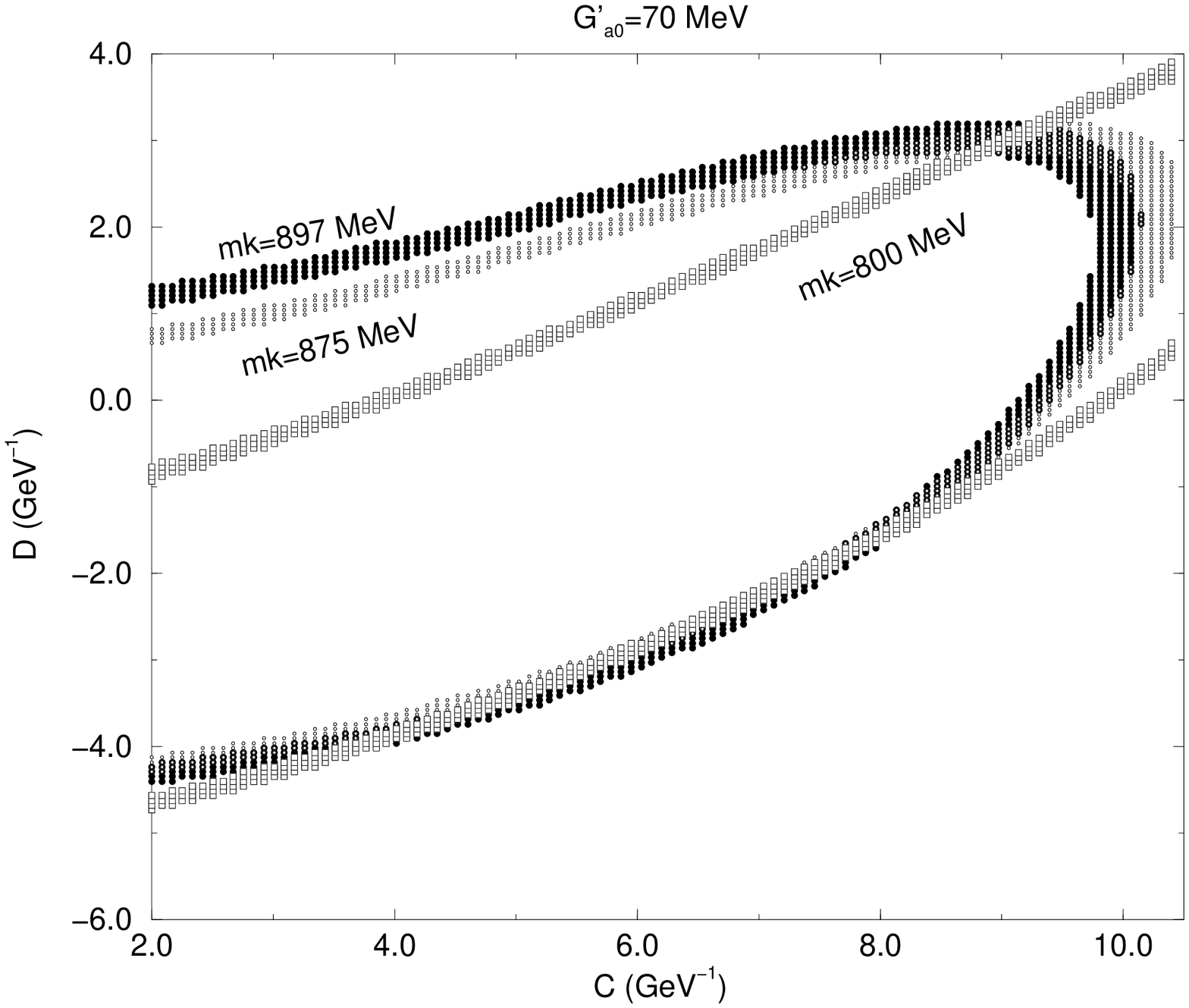, height=5in, angle=270}
\caption
{
The effect of $m_\kappa$ on the acceptable regions  consistent with the
$\eta'$ partial decay width. $G'_\sigma=370$ MeV and $G'_{f0}=64.6$ MeV.
}
\label{Fig_CD_mk}
\end{figure}

\begin{figure}
\centering
\epsfig{file=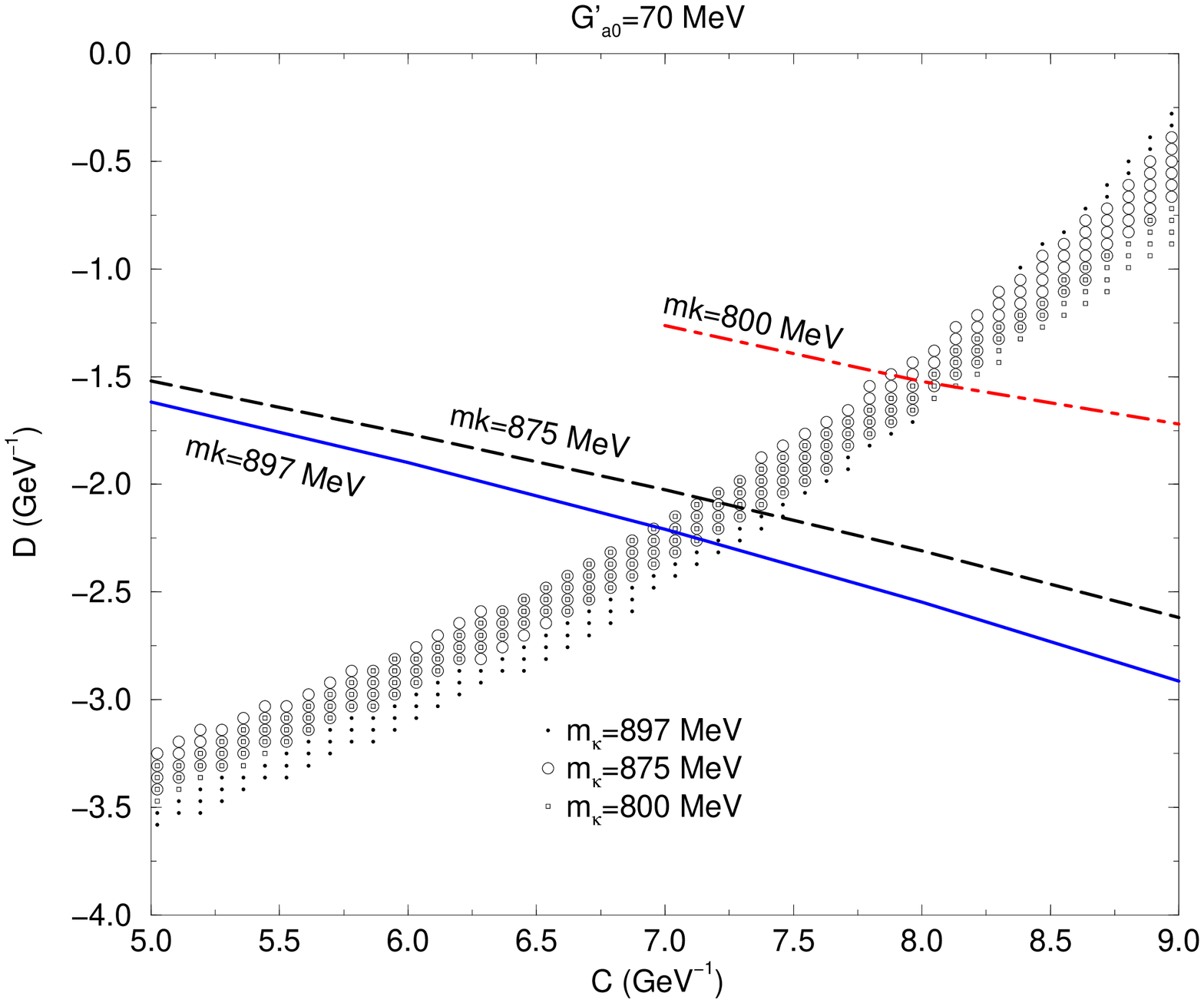, height=5in, angle=270}
\caption
{
Sensitivity of our computation to $m_\kappa$.
Strips represent regions consistent with the partial decay
width of $\eta'$, and lines represents the best least squared fits
of the normalized magnitudes of decay matrix element to the form
$\left| 1 +\alpha y \right|^2 + cx^2$ with ${\rm Re} \alpha=-0.0615$.
$G'_\sigma=370$ MeV and  $G'_{f0}=64.6$ MeV.
}
\label{Fig_alpha_mk}
\end{figure}

The value of the scalar mixing angle
$\theta_s\approx -21^{\circ}$ affects the entire calculation by its
presence in
the formulas[(\ref{gamma_kKp})-(\ref{gamma_feep})] relating the scalar
coupling constants
to the parameters $A$, $B$, $C$ and $D$.  Now $\theta_s$ is itself
determined by diagonalizing the isoscalar mass squared matrix obtained
from 
Eq.(\ref{L_mass}).  In this way, $\theta_s$ depends on the input value of 
$m_\kappa$.   The value $\theta_s\approx -20.33^o$ corresponds to 
$m_\kappa=897$ MeV but it was shown in \cite{Blk98b} that a range $865 MeV
< 
m_\kappa < 900$ MeV gave an acceptable description of $\pi K$ scattering.
Furthermore, reducing $m_\kappa$ to 800 MeV results in the ``ideal''
case where $\theta_s=0^o$.   In order to judge the sensitivity of our
results to changing $m_\kappa$
we repeat the present computation for two lower values 
$m_\kappa=875$ and  800 MeV.
As before  we scan the $CD$ plane for the acceptable regions
consistent with the $\eta' \rightarrow \eta \pi \pi$ decay width
(\ref{etap_exp}). We display the results in
Fig.\ref{Fig_CD_mk}
which shows that  the main effect of lowering $m_\kappa$ is in the 
$D>-1$ GeV$^{-1}$ region
, far from the physical region in which we extract $C$ and $D$.
We see in the same figure  that for $C\approx 4\rightarrow 8$ the effect
of changing $m_\kappa$ is negligible.  In Fig.\ref{Fig_alpha_mk} we have
displayed these regions together with the corresponding least squared
fits of the normalized magnitude of the decay matrix element of the
form (\ref{M_expand}).  As we see clearly in this figure, the value 
of $C$ extracted at the intersection of the lines with the strips changes
by a very small
amount as we go from $m_\kappa$=897 to 875 MeV.   On the other hand,  when
we go to the lower value of  $m_\kappa$=800 MeV, the 
the goodness of fit decreases  and in particular for $C<7$GeV$^{-1}$ we
get unacceptable fits.  Furthermore,  
for 
$m_\kappa =800$ MeV we get the partial decay width of 
$a_0(980)\rightarrow\pi\eta$ to be 124 MeV which is greater than the
total decay width and is inconsistent.
This agrees with the observation in \cite{Blk98b} 
that the values  $m_\kappa < 875$ MeV are not
favored. For the  value  $m_\kappa =$875  MeV the details of the
fit are given in the third column of Table \ref{T_I}.

\section{Summary and Discussion}

In this work, we studied in detail the $\eta'\rightarrow \eta 2\pi$
decay mode within the framework of a model in which the scalar meson
candidates $\sigma(560)$ (discussed in \cite{Har96}) and $\kappa(900)$
(discussed in \cite{Blk98a}) are combined into a nonet together with the
$f_0(980)$ and the $a_0(980)$.   The scalar mixing angle was calculated
\cite{Blk98b} in terms of these masses using Eq.(\ref{L_mass}) and the
various
scalar-pseudoscalar-pseudoscalar coupling constants were calculated in
terms of the parameters $A$, $B$, $C$ and $D$ in Eq.(\ref{L_NPP}).  In the
analysis of ref.\cite{Blk98b} the 
parameters $A$, $B$
and $\theta_s$ were found,  but parameters $C$ and $D$ were left
undetermined.  
As $\eta'$ decay probes these parameters, 
we have numerically searched this 
parameter space and found a unique  $C$ and
$D$ which  describes the experimental measurements on the partial decay
width of the $\eta'\rightarrow\eta\pi\pi$ as well as its energy
dependence.

Taking into account both the uncertainties in the scalar mixing angle
$\theta_s$ (as reflected in the value of $m_\kappa$) and in the
$\eta'\rightarrow\eta\pi\pi$ decay width we get for the scalar coupling
parameters
\begin{eqnarray}
A&=& 2.51 \rightarrow 2.87 \hskip .15cm {\rm GeV}^{-1}
\nonumber \\
B&=& -1.95 \rightarrow -2.34 \hskip .15cm {\rm GeV}^{-1}
\nonumber \\
C&=& 7.03 \rightarrow 7.39 \hskip .15cm {\rm GeV}^{-1}
\nonumber \\
D&=& -2.39 \rightarrow -1.95 \hskip .15cm {\rm GeV}^{-1}
\end{eqnarray}
These numbers are based on combining the second and third columns of
Table \ref{T_I}.  The  coupling constants relevant here are
listed in Table \ref{T_II}.

As a by-product of the present calculation we obtain an estimate of the
$a_0(980)$ width
\begin{equation}
\Gamma \left( a_0(980) \right) \approx 70 {\rm MeV}
\end{equation}
as discussed in Section III.  After this work was completed we found 
a very new experimental analysis \cite{Tei99} of the $\pi^- p \rightarrow
\eta\pi^+\pi^-$ and 
 $\pi^- p \rightarrow \eta\pi^0 n$ reactions which yields the same result
we have obtained from analysis of the $\eta'\rightarrow\eta\pi\pi$ decay.

\begin{figure}
\centering
\epsfig{file=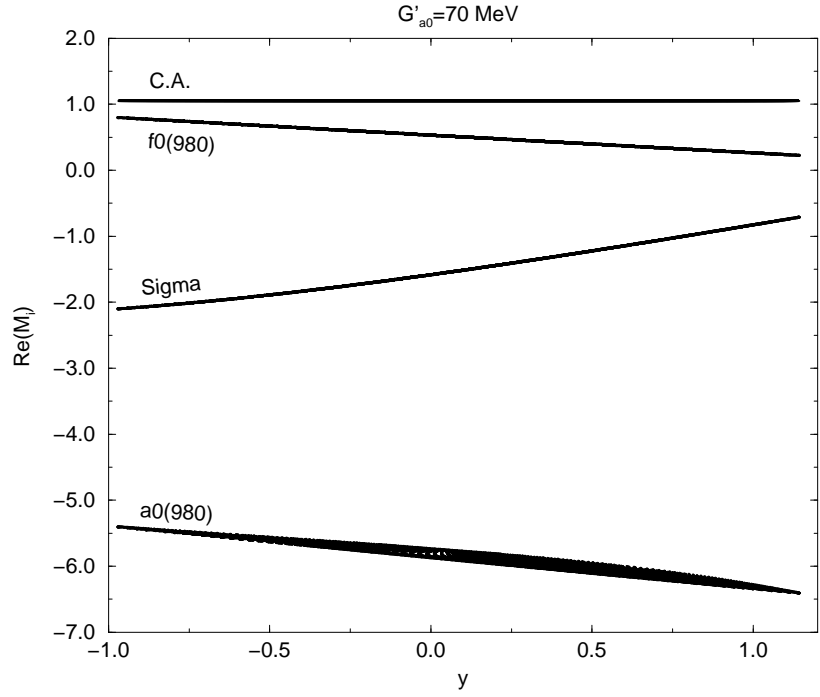, height=5in, angle=270}

\epsfig{file=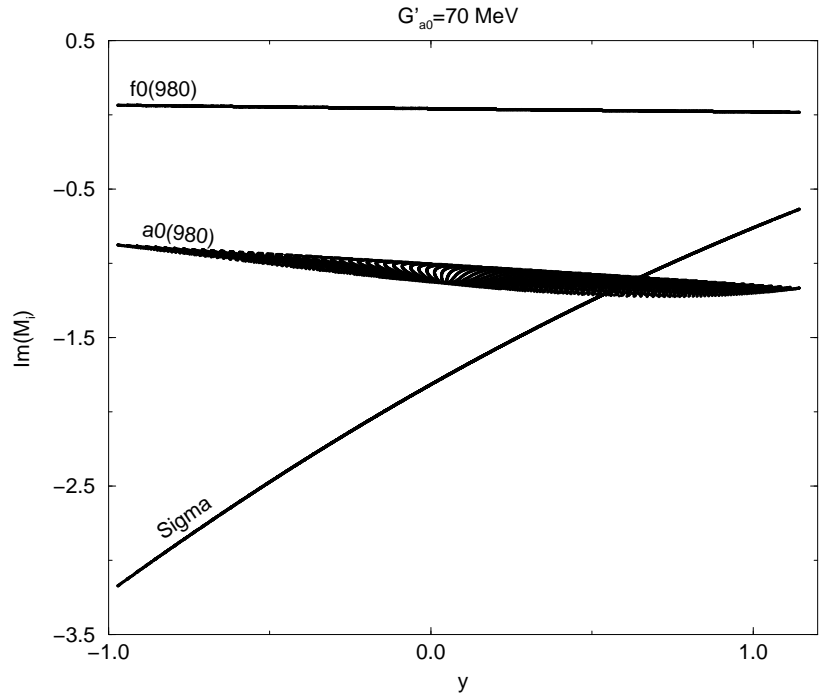, height=5in, angle=270}
\caption
{
Projections onto the Re $(M_i)-y$ and Im $(M_i)-y$ planes
of the individual scalar contributions \hskip .1cm to the 
decay matrix element corresponding to the result given in the second
column of Table \ref{T_I}.
}
\label{Fig_DP_Gpa70}
\end{figure}

It seems useful to ``dissect'' our model in order to get a qualitative
understanding of the $\eta'\rightarrow\eta\pi\pi$ process.  Thus
we have plotted, in Fig.\ref{Fig_DP_Gpa70}, the real and imaginary parts
of the
individual contributions of the terms in Eq.(\ref{M_individual}) to the
total decay
matrix element.  These figures again represent projections of the 
Re $M(x,y)$ and Im $M(x,y)$ surfaces onto the Re $M-y$ and Im $M-y$
planes;
the small $x$ dependences are thus visible as thickening of the curves.
First,
we observe that the ``current-algebra'' part of the amplitude, which
corresponds to the use of the minimal non-linear chiral Lagrangian of
pseudoscalar fields, is an order of magnitude too small to explain the
experimental result by itself.  On the other hand,  the $a_0(980)$
exchange contribution is clearly the main one for explaining the 
dominant real part of the amplitude.  Nevertheless the other contributions
are not negligible.  For example the cross term
2 [Re $M(\sigma)$] [Re $M(a_0)$] is of the same order as 
[Re $M(a_0)$]$^2$.  Furthermore the $\sigma$ meson exchange is seen to
give the largest contribution to Im $M$ for most of the  kinematical
range.

Note that we have used just the two input numbers, $C$ and $D$ (over and 
above the ones previously found) to satisfactorily fit the rate and energy
distribution of $\eta'\rightarrow\eta\pi\pi$.  Thus in the same framework,
with the same parameters, we are explaining $\pi\pi$ \cite{Har96} and $\pi
K$
\cite{Blk98a} scattering up to the 1 GeV range as well as
$\eta'\rightarrow
\eta\pi\pi$.
Our results may then be regarded as support for the correctness of both
the large $N_c$ approximation motivated approach to low energy dynamics
being employed as well as the effective Lagrangian model \cite{Blk98b} for
the
low lying scalar nonet outlined in the Introduction.
Of course, the ``microscopic'' structure of low lying scalars is an
interesting puzzle of present day particle physics which seems to require
a great deal of further experimental and theoretical work for its 
clarification.  For example, the study of radiative decays of the
$\phi(1020)$ is expected \cite{Ach89} to yield useful information.  As
discussed in more detail in \cite{Blk98b}, the value of the mixing angle
$\theta_s$,  about $-21^{\circ}$ and the mass spectrum   used here are 
what one would expect with a
somewhat distorted form of the $qq{\bar q} {\bar q}$ model \cite{Jaf77}.
A {\it priori},  however, our effective Lagrangian approach can
accommodate any microscopic model which yields a flavor nonet.

\acknowledgments

We are happy to thank Deirdre Black and Francesco Sannino for many helpful
discussions.  This work has been supported in part by DE-FG-02-92ER-40704.

\appendix

\section{}
Here we give, for convenience, the explicit form of the
scalar-pseudoscalar-pseudoscalar interaction \cite{Blk98b}.
Using isotopic spin invariance, the trilinear $N\phi\phi$ interaction 
from Eq. (\ref{L_NPP}) must have the form:

\begin{eqnarray}
-{\cal L}_{N\phi\phi} &=& \frac{\gamma_{\kappa K \pi}}{\sqrt 2} \left(
\partial_\mu {\bar K} \mbox{\boldmath ${\tau}$} \cdot \partial_\mu 
{\mbox{\boldmath ${\pi}$}}
\kappa + h.c. \right) + \frac{\gamma_{\sigma \pi \pi}}{\sqrt 2}
\sigma \partial_\mu \mbox{\boldmath ${\pi}$} \cdot \partial_\mu
{\mbox{\boldmath ${\pi}$}} \nonumber \\ 
&+& \frac{\gamma_{\sigma K  K}}{\sqrt 2}
\sigma \partial_\mu {\bar K} \partial_\mu {K}
+ \frac{\gamma_{f_0 \pi \pi}}{\sqrt 2}
f_0 \partial_\mu \mbox{\boldmath ${\pi}$} \cdot \partial_\mu 
 \mbox{\boldmath ${\pi}$}
+ \frac{\gamma_{f_0 K K}}{\sqrt 2}
f_0 \partial_\mu {\bar K} \partial_\mu {K} \nonumber \\
&+& \frac{\gamma_{a_0 K  K}}{\sqrt 2} \partial_\mu {\bar K} 
 \mbox{\boldmath ${\tau}$} \cdot {\bf a_0}
\partial_\mu {K} +
\gamma_{\kappa {K} \eta} \left(
{\bar \kappa} \partial_\mu K \partial_\mu {\eta} + h.c. \right) +
\gamma_{\kappa {K} \eta '} \left(
{\bar \kappa} \partial_\mu K \partial_\mu {\eta '} + h.c. \right)
\nonumber \\
&+& \gamma_{a_0 \pi\eta} {\bf a_0} \cdot \partial_\mu \mbox{\boldmath
${\pi}$} 
\partial_\mu \eta +
 \gamma_{a_0 \pi\eta'} {\bf a_0} \cdot \partial_\mu \mbox{\boldmath
${\pi}$} \partial_\mu \eta'  \nonumber \\
&+& \gamma_{\sigma \eta \eta} \sigma \partial_\mu \eta \partial_\mu \eta
+ \gamma_{\sigma \eta \eta'} \sigma \partial_\mu \eta \partial_\mu \eta'
+ \gamma_{\sigma \eta' \eta'} \sigma \partial_\mu \eta' \partial_\mu
\eta' \nonumber \\
&+&  \gamma_{f_0 \eta \eta} f_0 \partial_\mu \eta \partial_\mu \eta  
+  \gamma_{f_0 \eta \eta'} f_0 \partial_\mu \eta \partial_\mu \eta'  
+ \gamma_{f_0 \eta' \eta'} f_0 \partial_\mu \eta' \partial_\mu \eta',
\label{trilinear-interactions}
\end{eqnarray}
where the $\gamma$'s are the coupling constants.  The fields which appear
in this expression are the isomultiplets:
\begin{eqnarray}
K= \left( \begin{array}{c} K^+ \\ K^0 \end{array} \right),  
\quad {\bar K}= \left( \begin{array}{cc} K^- & {\bar K^0} \end{array}
\right)&,& \quad  
\kappa = \left( \begin{array}{c} \kappa ^+ \\ \kappa^0 \end{array}
\right), \quad
{\bar \kappa} = \left( \begin{array}{cc} {\kappa}^- & {\bar {\kappa}^0}
\end{array} \right),  \nonumber \\
\pi^\pm = \frac{1}{\sqrt 2} \left( \pi_1 \mp i\pi_2 \right)&,& \quad
\pi^0 = \pi_3,  \nonumber \\
a_0^\pm = \frac{1}{\sqrt 2} \left( a_{01} \mp ia_{02} \right)&,& \quad
a_0^0 = a_{03},
\label{isomultiplets}
\end{eqnarray}
in addition to the isosinglets $\sigma$, $f_0$, $\eta$ and $\eta'$.  The
$\gamma$'s are related to parameters $A$, $B$, $C$, $D$ 
of Eq. (\ref{L_NPP}) by
\begin{eqnarray}
\gamma_{\kappa  K \pi} & = & \gamma_{a_0 K K} = -2 A
\label{gamma_kKp} \\
\gamma_{\sigma \pi \pi} &=& 
2 B {\rm sin}\theta_s  - \sqrt{2} (B-A) {\rm cos }\theta_s 
\\
\gamma_{\sigma K {K}} &=& 
2 (2 B - A) {\rm sin}\theta_s -2 \sqrt{2} B {\rm cos }\theta_s  
\\
\gamma_{f_0 \pi \pi} &=& 
\sqrt{2} (A - B) {\rm sin }\theta_s - 2 B {\rm cos}\theta_s
\\
\gamma_{f_0 K {K}} &=& 
2 (A - 2 B) {\rm cos}\theta_s -2 \sqrt{2} B {\rm sin }\theta_s  
\label{f-K-barK}\\
\gamma_{\kappa {K} \eta} &=&  
C {\rm sin} \theta_p - \sqrt{2} (C-A) {\rm cos }\theta_p  
\\
\gamma_{\kappa {K} \eta '} &=&  
\sqrt{2} (A-C) {\rm sin }\theta_p -  
C {\rm cos} \theta_p  
\\
\gamma_{a_0 \pi \eta} &=&  
(C- 2A) {\rm sin }\theta_p -  
\sqrt{2} C {\rm cos} \theta_p  
\label{gamma_ape}
\\
\gamma_{a_0 \pi \eta'} &=&  
(2A - C) {\rm cos }\theta_p -  
\sqrt{2} C {\rm sin} \theta_p  
\\
\gamma_{\sigma\eta\eta} &=&
\left[ \sqrt{2}(B+D) -{1\over 2}(C + 2A + 4D) {\rm sin}2\theta_p
+ \sqrt{2}(C+D) {\rm cos}^2\theta_p \right] {\rm sin}\theta_s
\nonumber \\ &&
- 
\left[ (B+D) -{1\over \sqrt{2} }(C + 2 D) {\rm sin}2\theta_p
+ (A+D) {\rm cos}^2\theta_p  + C {\rm sin}^2\theta_p\right] 
{\rm cos}\theta_s
\\
\gamma_{\sigma\eta'\eta'} &=&
\left[ \sqrt{2}(B+D) + {1\over 2}(C + 2A + 4D) {\rm sin}2\theta_p
+ \sqrt{2}(C+D) {\rm sin}^2\theta_p \right] {\rm sin}\theta_s
\nonumber \\ &&
- 
\left[ (B+D) + {1\over \sqrt{2} }(C + 2 D) {\rm sin}2\theta_p
+ (A+D) {\rm sin}^2\theta_p  + C {\rm cos}^2\theta_p\right] 
{\rm cos}\theta_s
\\
\gamma_{\sigma\eta\eta'} &=&
\left[ \sqrt{2}(C+D){\rm sin}2\theta_p + (C + 2A + 4D) {\rm cos}2\theta_p
\right] {\rm sin}\theta_s
\nonumber \\ &&
- 
\left[ \sqrt{2} (C+2 D) {\rm cos}2 \theta_p + (A - C + D)
{\rm sin}2\theta_p \right] 
{\rm cos}\theta_s
\\
\gamma_{f_0\eta\eta} &=&
\left[- \sqrt{2}(B+D)+ {1\over 2}(C + 2A + 4D) {\rm sin}2\theta_p
- \sqrt{2}(C+D) {\rm cos}^2\theta_p \right] {\rm cos}\theta_s
\nonumber \\ &&
- 
\left[ (B+D) -{1\over \sqrt{2} }(C + 2 D) {\rm sin}2\theta_p
+ (A+D) {\rm cos}^2\theta_p  + C {\rm sin}^2\theta_p\right] 
{\rm sin}\theta_s
\\
\gamma_{f_0\eta'\eta'} &=&
-\left[ \sqrt{2}(B+D)+ {1\over 2}(C + 2A + 4D) {\rm sin}2\theta_p
+ \sqrt{2}(C+D) {\rm sin}^2\theta_p \right] {\rm cos}\theta_s
\nonumber \\ &&
- 
\left[ (B+D) +{1\over \sqrt{2} }(C + 2 D) {\rm sin}2\theta_p
+ (A+D) {\rm sin}^2\theta_p  + C {\rm cos}^2\theta_p\right] 
{\rm sin}\theta_s
\\
\gamma_{f_0\eta\eta'} &=&
-\left[ \sqrt{2}(C+D){\rm sin}2\theta_p + (C + 2A + 4D) {\rm
cos}2\theta_p
\right] {\rm cos}\theta_s
\nonumber \\ &&
- 
\left[ \sqrt{2} (C+2 D) {\rm cos}2 \theta_p + (A - C + D)
{\rm sin}2\theta_p \right] 
{\rm sin}\theta_s
\label{gamma_feep}
\end{eqnarray}
where $\theta_s$ is the scalar mixing angle defined in Eq.(\ref{S_mix})
while
$\theta_p$ is the pseudoscalar mixing angle defined by
\begin{equation}
\left( 
\begin{array}{c} 
        \eta\\ 
        \eta' 
\end{array} 
\right) =
\left( 
\begin{array}{c c} 
{\rm cos} \theta_p  & -{\rm sin} \theta_p \\
{\rm sin} \theta_p  &  {\rm cos} \theta_p 
\end{array} 
\right)
\left( 
\begin{array}{c} 
 (\phi^1_1+\phi^2_2)/ \sqrt{2} \\ \phi^3_3 
\end{array} 
\right),
\label{eta-etap}
\end{equation}
where $\eta$ and $\eta'$ are the fields which diagonalize the pseudoscalar
squared mass matrix.  We adopt here the conventional value $\theta_p
\approx 37^o$.  (see \cite{Blk98b} for additional discussion.)

\begin{table}
\begin{center}
$\begin{array}{|c|c|c|c|}
 \hline 
     \gamma_{\sigma \pi\pi}     &    7.27   &   7.27  &  8.36 
\\ \hline  
     \gamma_{\sigma K K}        &    9.63   &   9.63  &  10.44
\\ \hline
     \gamma_{\sigma\eta\eta}    &    3.90   &   4.11  &  4.30
\\ \hline
     \gamma_{\sigma\eta\eta'}   &   1.25    &   2.65  &  2.61
\\ \hline
     \gamma_{\sigma\eta'\eta'}  &   -3.82   &  -1.43  &  -2.09
\\ \hline
     \gamma_{f\pi\pi}           &   1.47    &   1.47  &  2.53
\\ \hline
     \gamma_{fK K}              &   10.11   &  10.11  &  12.76
\\ \hline
     \gamma_{f\eta\eta}         &   1.50    &  1.72   &  2.78
\\ \hline
     \gamma_{f\eta\eta'}        &   -10.19  &  -9.01  &  -9.34
\\ \hline
     \gamma_{f\eta'\eta'}       &   1.04    &  2.60   &  2.04
\\ \hline
     \gamma_{a\pi\eta}          &    -6.87  &  -6.80  &  -7.28
\\ \hline   
     \gamma_{a\pi\eta'}         &    -8.02  &  -7.80  &  -7.38
\\ \hline   
\end{array}$
\end{center}  
\caption{
Predicted coupling constants corresponding to the columns in Table
\ref{T_I}.  All units are in GeV$^{-1}$.}
\label{T_II}
\end{table}


\begin{thebibliography}{99}
\bibitem{Torn95} See, for example, {N.A.~T\"ornqvist}, Z. Phys. 
{\bf C68}, 647 (1995) and references therein.  In addition see
{N.A.~T\"ornqvist} and M. Roos, Phys. Rev. Lett. {\bf 76}, 1575
(1996).

\bibitem{Ish96} S. Ishida, M.Y. Ishida, H. Takahashi, T. Ishida,
K. Takamatsu and T Tsuru, Prog. Theor. Phys. {\bf 95}, 745 (1996).

\bibitem{Mor93} 
D. Morgan and M. Pennington, Phys. Rev. {\bf D48},  1185  (1993).

\bibitem{Jans95}
{G.~Janssen, B.C.~Pearce, K.~Holinde and J.~Speth}, Phys. Rev. {\bf  
D52},  2690  (1995).

\bibitem{Bol93} A.A. Bolokhov, A.N. Manashov, M.V. Polyakov and
V.V. Vereshagin, Phys. Rev. {\bf D48}, 3090 (1993).  See also
V.A. Andrianov and A.N. Manashov, Mod. Phys. Lett. {\bf A8}, 2199
(1993).  Extension of this string-like approach to the $\pi K$ case
has been made in V.V. Vereshagin, Phys. Rev. {\bf D55}, 5349 (1997)
and very recently in A.V. Vereshagin and V.V. Vereshagin hep-ph/9807399,
which is consistent with a light $\kappa$ state.

\bibitem{Ach94} N.N. Achasov and G.N. Shestakov, Phys. Rev. {\bf
D49}, 5779 (1994).

\bibitem{Kam94}{R. Kam\'inski}, {L. Le\'sniak} and J. P. Maillet,
Phys. Rev. {\bf D50}, 3145 (1994).

\bibitem{Sv96} M. Svec, Phys. Rev. {\bf D53}, 2343 (1996). 

\bibitem{van86} E. van Beveren, T.A. Rijken, K. Metzger,
C. Dullemond, G. Rupp and J.E. Ribeiro, Z. Phys. {\bf C30}, 615
(1986). E. van Beveren and G. Rupp, hep-ph/9806246, 248.  See also
J.J. de Swart, P.M.M. Maessen and T.A. Rijken, U.S./Japan Seminar on the 
YN Interaction, Maui, 1993 [Nijmegen report THEF-NYM 9403].

\bibitem{Del92} R. Delbourgo and M.D. Scadron, Mod. Phys. Lett. {\bf
A10}, 251 (1995).  See also D. Atkinson, M. Harada and A.I. Sanda,
Phys.~Rev. {\bf D46}, 3884 (1992).

\bibitem{Ol98} J.A. Oller, E. Oset and J.R. Pelaez, hep-ph/9804209

\bibitem{Ish98}
S.~Ishida, M.~Ishida, T.~Ishida, K.~Takamatsu and T.~Tsuru,
Prog. Theor. Phys. {\bf 98}, 621 (1997). See also M. Ishida and S. Ishida,
Talk given at 7th International Conference on Hadron Spectroscopy (Hadron
97), Upton, NY, 25-30 Aug. 1997, hep-ph/9712231. 

\bibitem{Torn97} N.A. T\"ornqvist, hep-ph/9711483, hep-ph/9712479.

\bibitem{Anis97}A.V. Anisovich and A.V. Sarantsev, Phys. Lett. {\bf B413},
137 (1997).

\bibitem{Eli98}V. Elias, A.H. Fariborz, Fang Shi and T.G. Steele,
Nucl. Phys. {\bf A633}, 279 (1998).

\bibitem{Dmi96} V. Dmitrasinovi\'c, Phys. Rev. {\bf C53}, 1383 (1996).

\bibitem{San95}
F. Sannino and J. Schechter, Phys. Rev.  {\bf D52},  96  (1995).

\bibitem{Har96}
M. Harada, F. Sannino and J. Schechter, Phys. Rev. {\bf D54},  
1991 (1996), Phys. Rev. Lett. {\bf 78}, 1603 (1997).


\bibitem{Blk98a}D. Black, A.H. Fariborz, F. Sannino, and J. Schechter,
Phys. Rev. {\bf D58}, 054012 (1998). 

\bibitem{Ast88}D. Aston {\it et al}, Nucl. Phys {\bf B296}, 493 (1988). 


\bibitem{Blk98b}D. Black, A.H. Fariborz, F. Sannino, and J. Schechter,
hep-ph/9808415; Phys. Rev. {\bf D}, to be published.

\bibitem{Oku63}S. Okubo, Phys. Lett. {\bf 5}, 165 (1963). See also
G. Zweig, CERN report 8182/TH 40/ and 8419/TH 412 (1964); J. Iizuka,
Prog. Theor. Phys. Suppl 37-{\bf 8}, 21 (1966).

\bibitem{Cal69}C. Callan, S. Coleman, J. Wess and B. Zumino,
Phys. Rev. {\bf 177}, 2247 (1969). 

\bibitem{Jaf77}R.L. Jaffe, Phys. Rev. {\bf D15}, 267 (1977);
R.L. Jaffe and F.E. Low, Phys. Rev. {\bf D19}, 2105 (1979).

\bibitem{Isg90}N. Isgur and J. Weinstein, Phys. Rev. {\bf D41}, 2236
(1990).

\bibitem{Cro67} J. Cronin, Phys. Rev. {\bf 161}, 1483 (1967).

\bibitem{Schw68}J. Schwinger, Phys. Rev. {\bf 167}, 1432 (1968).

\bibitem{Maj68} D. Majumdar, Phys. Rev. Lett.{\bf 21}, 502 (1968).

\bibitem{DiV81} P. DiVecchia, F. Nicodemi, R. Pettorino and G.Veneziano,
Nucl. Phys. {\bf B181}, 318 (1981).

\bibitem{Sch73} J. Schechter and Y.Ueda, Phys. Rev. {\bf D3}, 2874 (1971);
see also E.{\bf D8}, 987 (1973).

\bibitem{Sing75}C.Singh and J.Pasupathy, Phys. Rev. Lett. {\bf 35}, 1193
(1975).

\bibitem{Desh78} N. Deshpande and T.Truong  , Phys. Rev. Lett. {\bf 41},
1579
(1978).

\bibitem{Bra80} A. Braman and E. Masso, Phys. Lett. {\bf 93B}, 65 (1980).

\bibitem{Wein79} S. Weinberg, Physica {\bf 96A}, 327 (1979). J. Gasser
and H.  Leutwyler, Ann. of Phys. {\bf 158},  142 (1984); J. Gasser and
H. Leutwyler, Nucl. Phys. {\bf B250},  465 (1985). A recent review is  
given by Ulf-G. Mei\ss ner, Rept. Prog. Phys. {\bf 56}, 903 (1993).

\bibitem{PDG}Review of Particle Physics, Euro. Phys. J {\bf C3} (1999).


\bibitem{Ald86} D. Alde, et.al., Phys. Lett. {\bf B177}, 115 (1986).

\bibitem{Tei99}S. Teige et al, Phys. Rev. {\bf D54}, 012001 (1999).

\bibitem{Ach89} 
N. Achasov and V. Ivanchenko,  Nucl. Phys. {\bf B315}, 465 (1989);
R. Akhmetshin et al, Phys. Lett. {\bf B415}, 452 (1997);
M. Achasov et al, hep-ex/9807016;
F. Close, N. Isgur and S. Kumano, Nucl. Phys. {\bf B389}, 513 (1993).

\end{thebibliography}
\end{document}